\documentclass[3p,review,sort&compress]{elsarticle}
\usepackage{endfloat}
\usepackage{titletoc}

\usepackage{graphicx}
\usepackage{amsmath}
\usepackage{amssymb}
\usepackage{hyperref}
\newcommand{\iddots}{\reflectbox{$\ddots$}}

\begin{document}

\begin{frontmatter}
\title{Elastic constants and mechanical properties of PEDOT from first principles calculations}

\author[phys]{R. O. Agbaoye}
\ead{agbaoye@physics.unaab.edu.ng}

\author[physcs]{P. O. Adebambo}
\ead{adebambo@physics.unaab.edu.ng}

\author[phys]{J. O. Akinlami}
\ead{johnsonak2000@yahoo.co.uk}

\author[chem]{T. A. Afolabi}
\ead{afolabita@funaab.edu.ng}

\author[solar]{Smagul Zh. Karazhanov}
\ead{smagul.karazhanov@ife.no}

\author[istm]{Davide Ceresoli\corref{cor1}}
\ead{davide.ceresoli@cnr.it}

\author[phys]{G. A. Adebayo\corref{cor1}}
\ead{adebayo@physics.unaab.edu.ng}

\address[phys]{Department of Physics, Federal University of Agriculture, PMB 2240, Abeokuta, Nigeria}
\address[physcs]{Department of Physical and Computer Sciences, McPherson University, Abeokuta, Nigeria}
\address[chem]{Department of Chemistry, Federal University of Agriculture, PMB 2240, Abeokuta, Nigeria}
\address[solar]{Department for Solar Energy, Institute for Energy Technology, P.O Box 40, NO 2027-Kjeller, Norway}
\address[istm]{CNR-ISTM and INSTM, c/o Dipartimento di Chimica, Universit\`a degli studi di Milano,
via Golgi 19, 20133 Milano, Italy}

\cortext[cor1]{Corresponding author}

\begin{abstract}
In this work, we report about the electronic and elastic properties
of crystalline poly(3,4-ethylene\-dioxy\-thiophene), known as PEDOT,
in an undiluted state, studied in the framework of semilocal DFT,
using the PBE and PBEsol exchange-correlation functional and PAW
pseudopotentials. Contrary to previous molecular dynamics simulations,
our calculations revealed that the most stable state structure of
pristine PEDOT is monoclinic. We calculated the 13 independent elastic
constants and the elastic compliance which enables us to establish
other elastic properties of pristine PEDOT; the Pugh's ratio and the
Vicker's hardness computed with PBE and PBEsol are in good agreement
with each other. Finally, we compute the directional elastic modulii
and found remarkable differences between different DFT functionals.
\end{abstract}


\end{frontmatter}

\section{Introduction}\label{sec:intro}
As an organic semiconducting polymer, poly(3,4-ethelyndioxythiopene)
(PEDOT) is finding highly significant applications in modern
day technology. PEDOT, when doped with other polymers/materials,
provides the ground to develop novel functional materials. Due to
its flexible nature, thin films of PEDOT:PSS (PSS~=~polystyrene sulfonate)
are being employed in flexible electronic devices~\cite{Zhou2014}.
Since the first synthesis of PEDOT~\cite{Cho2014}, several experimental
studies have been carried out on doped and pristine either as thin-film or
bulk material. These studies range from synthesis, doping, chemical preparation,
thermoelectric properties to the electronic, optical and structural
determination.~\cite{Zhou2014,Cho2014,Cho2015,Pyshkina2010,Pei1994,%
TranVan2001,Aasmundtveit1999} The conductivity, atmospheric stability, band gap,
thermoelectric figure of merit among other properties of PEDOT have improved
over time, but the elastic and thermodynamics properties have neither been
studied nor reported until now.~\cite{Shi2015,Zhang2015}.

As mentioned earlier, PEDOT is an optically active conductive
conjugated polymer~\cite{Cho2015}. It exists as an organic
semiconductor with a small direct band gap in the pure
state; with exceptional environmental stability and electrical
conductivity~\cite{Pyshkina2010,Pei1994,TranVan2001,Aasmundtveit1999}.
Synthesis in the pristine state via chemical polymerization results
in blue-black color, while it becomes to almost transparent when
doped~\cite{Pei1994,TranVan2001}.  This is due to the ability to
change from benzoic shape to quinoid structure when doped with PSS
and from aromatic-like structure to quinoid-like structure when
doped with tosylate~\cite{Lenz2011,Shi2015,Kim2008,Zhang2015}.
In pristine, undoped PEDOT, the gap between its highest occupied
molecular orbital (HOMO) and its lowest unoccupied molecular
orbitals (LUMO) is reported as 1.5~eV~\cite{TranVan2001}, 1.6--1.7~eV~\cite{Aasmundtveit1999}
and 1.64~eV~\cite{Havinga1996}. On the other hand, Refs.~\cite{ Shi2015,Kim2008,Zhang2015}
report band gaps of 0.37~eV with the B3LYP exchange-correlation functional,
0.45~eV with PBE, 0.16~eV with PBE-D and 0.53~eV with the HSE06 hybrid functional.
The discrepancy between the experimental and theoretical band gap might
be related to the fact that experimental measurements are carried on
dispersed PEDOT in thin films, while the theoretical calculations address
only the pure crystalline phases.

These important attributes of PEDOT make it an important material to
be studied both theoretically and experimentally. Recent calculations
by Wen Shi et al.~\cite{Shi2015} provided electronic and thermoelectric
properties of both pristine and doped PEDOT. A theoretical investigation
of the mechanical properties and how they influence the electrical
conductivity of PEDOT, will shed more light on the efficiency and
reliability of PEDOT for flexible-electronic applications. As a first
step towards elucidating this, we calculated the elastic constants and
mechanical properties of PEDOT, and thermodynamic properties (Debye
temperature, specific heat capacity) by first principles DFT lattice
dynamics calculations.

To compute elastic properties, in principle one has to deal with 81
independent elastic constants associated with a crystal. However, using
the approach employed by Newman, these are reduced to 21 for all crystal
structure~\cite{Newnham2005}; the symmetry operations of a monoclinic
crystal contribute to reducing the 21 independent elastic constants
down to 13. Using the Voigt notation, the elastic constants tensor for
a monoclinic $b$ unique axis crystal can be written as:

\begin{equation}
C = \begin{bmatrix}
C_{11} & C_{12} & C_{13} & 0      & C_{15} & 0      \\
       & C_{22} & C_{23} & 0      & C_{25} & 0      \\
       &        & C_{33} & 0      & C_{35} & 0      \\
\vdots &        &        & C_{44} & 0      & C_{46} \\
       & \iddots&        &        & C_{55} & 0      \\
       &        & \hdots &        &        & C_{66} \\
\end{bmatrix},
\end{equation}
where the dots mean that the matrix is symmetric.

From the elastic constants, one can determine various elastic moduli:
the shear modulus, Young modulus, the Poisson ratio, and the bulk
modulus~\cite{Newnham2005,Greaves2011,Li2011}. The Poisson ratio gives
information about the type of bond that is possess in any material
which later predicts the properties of such material when a load is
exerted on it~\cite{Greaves2011}. A Poisson ratio within the range
0.25--0.42 characterizes a material dominated by metallic bonding~\cite{Li2011}.
In glasses, ceramics and semiconductors, the Poisson ratio is close to 0.25.
However, Greaves and co-workers predicted a Poisson ratio approximately 0.33
in polymers~\cite{Greaves2011}. Therefore, understanding how PEDOT is
classified, in terms of mechanical properties, is extremely interesting.

\section{Computational methods}\label{sec:computational}
Experimentally, PEDOT is semicrystalline and can be obtained with
different degrees of crystallinity. Despite the fact that X-ray
diffraction experiments~\cite{Aasmundtveit1999,TranVan2001} display
sharp peaks, corresponding to the lamellar structure of the polymer,
the crystalline structure of undoped PEDOT has not been fully determined
(only lattice parameters assuming an orthorhombic structure have been
reported).

\begin{figure}
\begin{center}
\includegraphics[width=0.35\columnwidth]{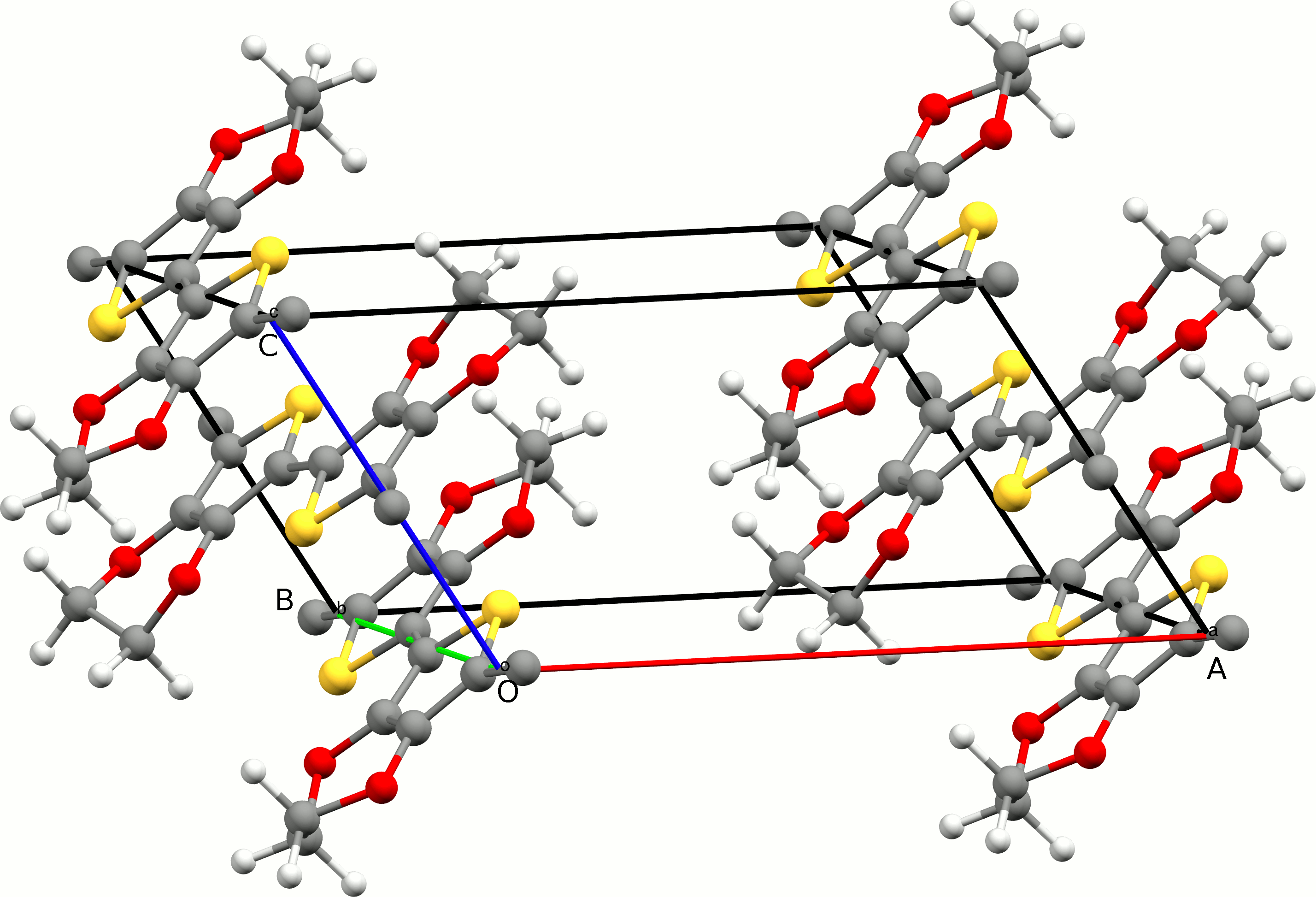}
\includegraphics[width=0.25\columnwidth]{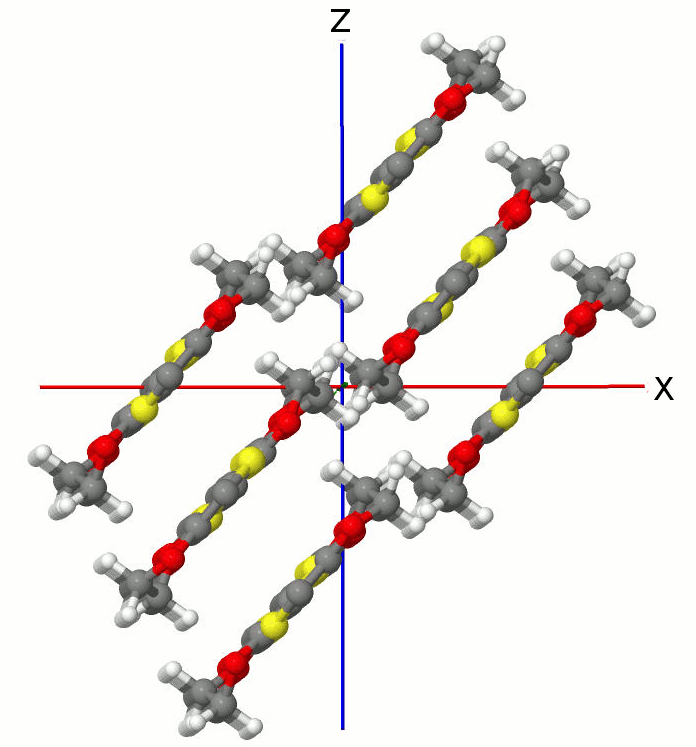}
\includegraphics[width=0.3\columnwidth]{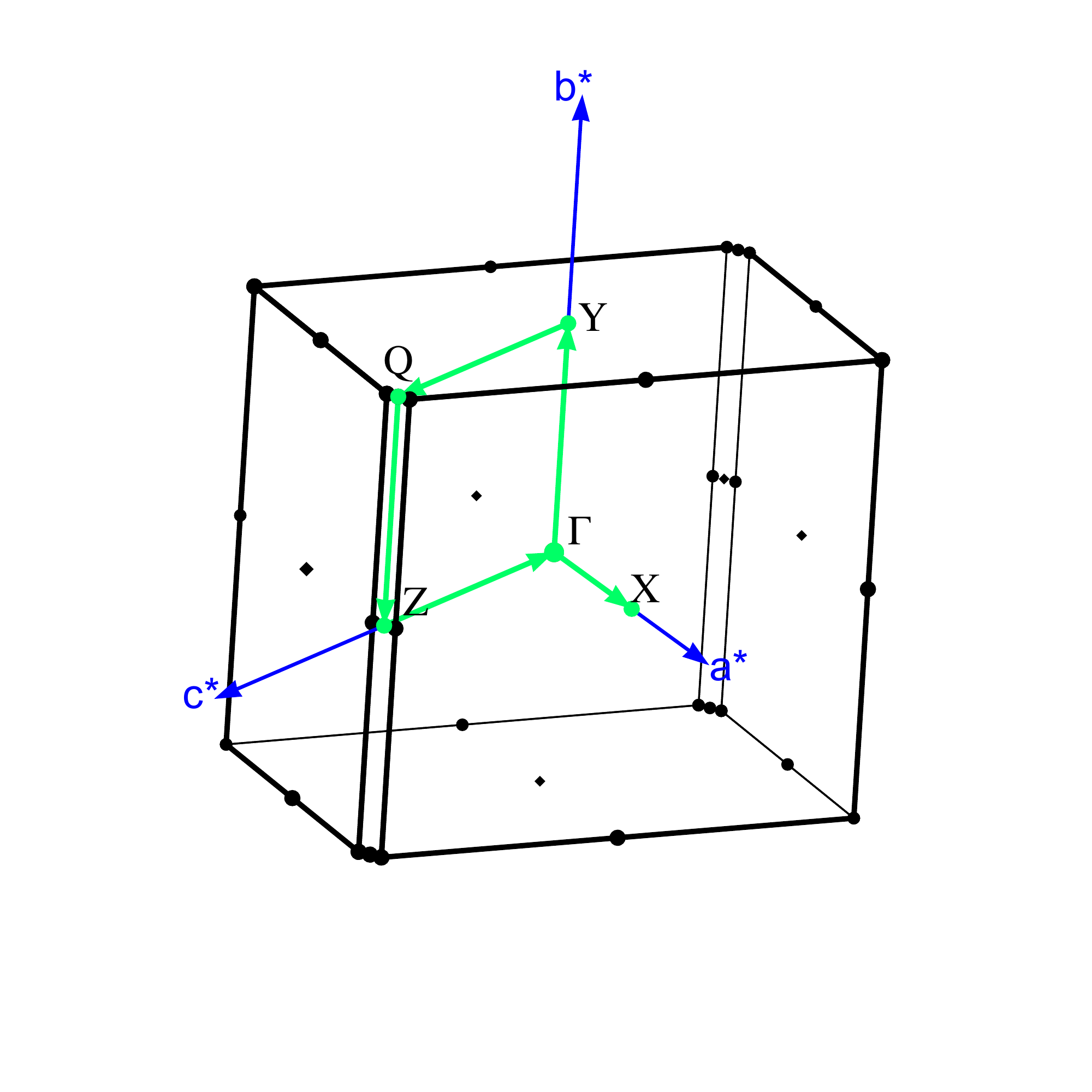}
\caption{(Left) Monoclinic PEDOT crystal structure optimized at the
PBEsol+D2 level.  Carbon atoms are grey, oxygen red, sufur yellow,
hydrogen white. The crystallographic axes $a$, $b$, and $c$ are
marked in red, green and blue, respectively.
(Center) View along the polymer chain. (Right) Brillouine
zone and reciprocal point path for band structure.}
\label{fig:cell}
\end{center}
\end{figure}

Previous DFT calculations, have shown that the crystal structure of
PEDOT can be either orthorhombic~\cite{Kim2008,Lenz2011,Zhang2015},
or monoclinic~\cite{Shi2015}. In the present work, we started from the
monoclinic structure of Ref.~\cite{Shi2015} (space group \emph{P2/c})
of pristine-type PEDOT, and performed full relaxation of the lattice.

We used two semilocal exchange and correlation functionals, the
PBE~\cite{Perdew1996} and PBEsol~\cite{Perdew2008}. In PBE calculations
we used norm-conserving pseudopotentials\footnote{C.pbe-nc.UPF,
O.pbe-nc.UPF, S.pbe-n-nc.UPF, H.pbe-n-nc.UPF} and a plane
wave cutoff of 120~Ry. In PBEsol calculations we used PAW
pseudopotentials\footnote{C.pbesol-n-kjpaw\_psl.0.1.UPF,
O.pbesol-n-kjpaw\_psl.0.1.UPF, S.pbesol-n-kjpaw\_psl.0.1.UPF,
H.pbesol-kjpaw\_psl.0.1.UPF}, from \textsl{PSlibrary}~\cite{PSLIBRARY}
and 75~Ry (750~Ry) energy cutoff for the planewaves (density). Together
with a Monkhorst-Pack k-points grid of 4$\times$7$\times$8 points, this
setup leads to well converged total energy within 0.8~meV/atom. The
k-points grid was chosen in order to sample the reciprocal space most
uniformly possible, along the three reciprocal lattice vectors.

The optimization calculations were performed for both monoclinic and
orthorhombic crystal structures, these calculations revealed the most
stable phase of monoclinic pristine PEDOT. We initially performed the
variable-cell relaxation with a smaller k-point mesh (2$\times$4$\times$4)
and then, to reduce the force on the ions and to obtain the most
stable state at minimum energy, the angle $\beta$ between the lattice
parameters $a$ and $c$ was optimized to achieve a global minimum state of
the crystal using the finer 4$\times$7$\times$8 kpoint grid. Finally,
we performed a self-consistent field (SCF) single point calculation with a
6$\times$10$\times$11 k-point mesh and calculated the electronic band
structure along the high symmetry lines of the Brillouin Zone. To obtain
a smooth density of states (DOS), we performed non-SCF with a denser
Monkhorst-Pack grid of 7$\times$11$\times$12.

We calculated the elastic constants of the molecular PEDOT by exerting
strains that cause either longitudinal, transverse or both longitudinal
and transverse distortion to the system, thereby calculating the
stress that takes it back to its initial configuration, the strain
and the obtained stress are then fitted to get the 13 independent
elastic constants. Since the energy derivatives are more sensitive
to the convergence parameters, we calculated the elastic constants
with a 2$\times$4$\times$4, 3$\times$5$\times$5, 4$\times$7$\times$8,
5$\times$8$\times$9, 6$\times$9$\times$10 and 7$\times$11$\times$12
Monkhorst-Pack grid. These correspond to 18, 24, 72, 115, 204, 258
k-points and 20, 38, 114, 181, 332, 463 k-points for PBE and PBEsol.

The elastic constants are calculated using the finite difference
approach~\cite{ELASTIC}. The small strain $\epsilon_j$ applied to perturb
a crystal, and the stress tensor $\alpha_j$ that tends to return it to
equilibrium, are related by:

\begin{equation}
  \alpha_j = \sum_{j=1}^6 C_{ij}\, \epsilon_i
\end{equation}

We perturbed the crystal by a set of 3$\times$3 strain tensors
which varies the length of lattice parameters, the size of the
crystal along the $yz$, $xz$, $xy$ planes, for a small applied strain
($\epsilon$) of magnitude $-$0.0075, $-$0.0025, 0.0025 and 0.0075. Then,
elastic constants are calculated [16-18, 20] from Voigt-Heuss-Hill
approximation. Then, we calculated bulk modulus, Young modulus,
shear modulus and the Poisson ratio for PEDOT using the approach
described in Refs.~\cite{Li2011,DalCorso2016,Shahsavari2009}. Other
calculated properties are the Pugh's modulus ratio and Vicker's
hardness~\cite{Zhou2014,Guler2015,Tian2012,Lechner2016}. This method,
although computationally expensive, has proven to be very accurate method
of determining the elastic properties of both organic and inorganic
crystals~\cite{Yao2007} with respect to other methods~\cite{Guler2015}.
Finally, we calculated the dependence of the elastic modulii on the
spatial direction, according to Ref.~\cite{Gaillac2016}.

\section{Results and discusssion}\label{sec:results}
The lattice parameters $a$, $b$, $c$ and monoclinic angle $\beta$
are reported in Tab.~\ref{tab:lattice}, together with experimental
data and previous calculations. The orthorhombic
structures can be seen as monoclinic by the following transformation:
\begin{equation}
\vec{a'} = \vec{a}+\vec{c}, \quad \vec{b'} = \vec{b},
\quad \vec{c'} = \vec{c},
\end{equation}
where $a'$, $b'$ and $c'$ are lattice spacings of the orthorhombic
structure. In Tab.~\ref{tab:lattice}, for sake of comparison, for the
orthorhombic structures, we report also the corresponding monoclinic
lattice parameters, as well as the crystal cell volume.  The $b$-unique
monoclinic crystal structure of pristine PEDOT~\cite{Shi2015} consists of
four units of ethylene dioxythiophene (EDOT), arranged into two parallel
polymeric chains as illustrated in Fig.~\ref{fig:cell}.

\begin{table*}
\begin{center}\begin{tabular}{l|lll|lllll|l}
\hline\hline
Method (XC functional) & $a'$ (\AA) & $b'$ (\AA) & $c'$ (\AA)
& $a$ (\AA) & $b$ (\AA) & $c$ (\AA) & $\beta (^\circ)$ & Volume (\AA$^3$) & Ref.\\
\hline
PW-NC (PBE)     & $-$   & $-$   & $-$ & 10.843 & 7.878 & 7.465 & 124.1 & 556.08 & This work \\
PW-PAW (PBEsol) & $-$   & $-$   & $-$ & 10.830 & 7.836 & 8.053 & 124.9 & 560.78 & This work \\
PW-NC (PBE+D2)  & $-$   & $-$   & $-$ & 10.773 & 7.878 & 7.445 & 123.0 & 534.36 & This work \\
PW-PAW (PBEsol+D2) & $-$   & $-$   & $-$ & 10.173 & 7.826 & 7.438 & 122.3 & 501.22 & This work \\
\hline
PW-USPP (PW91)  & 11.8  & 7.8  & 6.9  & 13.669 & 7.8   & 6.9   & 120.3 & 635.08 & \cite{Lenz2011} \\
PW-NCPP (PBE)   & 10.52 & 7.935 & 7.6 & 12.978 & 7.935 & 7.6   & 125.8 & 634.42 & \cite{Kim2008,Zhang2015} \\
PW-PAW (PBE+D2) & $-$   & $-$   & $-$ & 12.000 & 7.820 & 7.040 & 123.0 & 554.05 & \cite{Shi2015} \\
\hline
Experiment      & 14.0  & 7.8  & 6.8  & 15.556 & 7.8   & 6.8   & 115.9 & 742.56 & \cite{Aasmundtveit1999} \\
Experiment      & 10.52 & 7.87 & 5.66 & 12.052 & 7.87  & 5.88  & 119.2 & 468.60 & \cite{TranVan2001} \\
\hline\hline
\end{tabular}\end{center}
\caption{Optimized lattice parameters of monoclinic PEDOT compared with
experiments and other theoretical computation. $a'$, $b'$ and $c'$ are the
lattice parameters of the orthorhombic structure. $a$, $b$, $c$ and $\beta$
are the lattice parameters of the monoclinic structure. PW=plane wave; NCPP=norm
conserving pseudopotentials; USPP=ultrasoft pseudopotentials; PAW=projector augmented wave.}
\label{tab:lattice}
\end{table*}

Despite being in the reported experimental range, the optimized lattice
parameters using the PBE and PBEsol exchange-correlation functionals
overestimates the volume of the system.  This is to be expected, since
the PBE family is known to under-bind a large class of systems, molecular
and periodic. Even if PBEsol was designed specifically to reproduce the
lattice spacing of inorganic systems, it is still unable to describe the
weak dispersion forces that characterize molecular and polymeric crystals.
For instance, polyethylene crystal was predicted to be unbound using the
semilocal BLYP functional and it could be stabilized by the inclusion
of an empirical term.~\cite{Serra2000} From Tab.~\ref{tab:lattice} it
is evident that van der Waals and dispersion forces are essential to
describe the cohesive energy of crystalline PEDOT. Therefore we employed
the empirical correction ($+$D2) proposed by Grimme~\cite{Grimme2011}.
The $b$ lattice parameter (parallel to the polymer chains, where chemical
bonding is covalent) is well described both by PBE and PBEsol, with and
without vdW corrections. Notably, the difference between the PBE+D2 and
PBEsol+D2 equilibrium volume is $\sim$6\%.

\begin{figure}
\begin{center}
\includegraphics[width=\columnwidth]{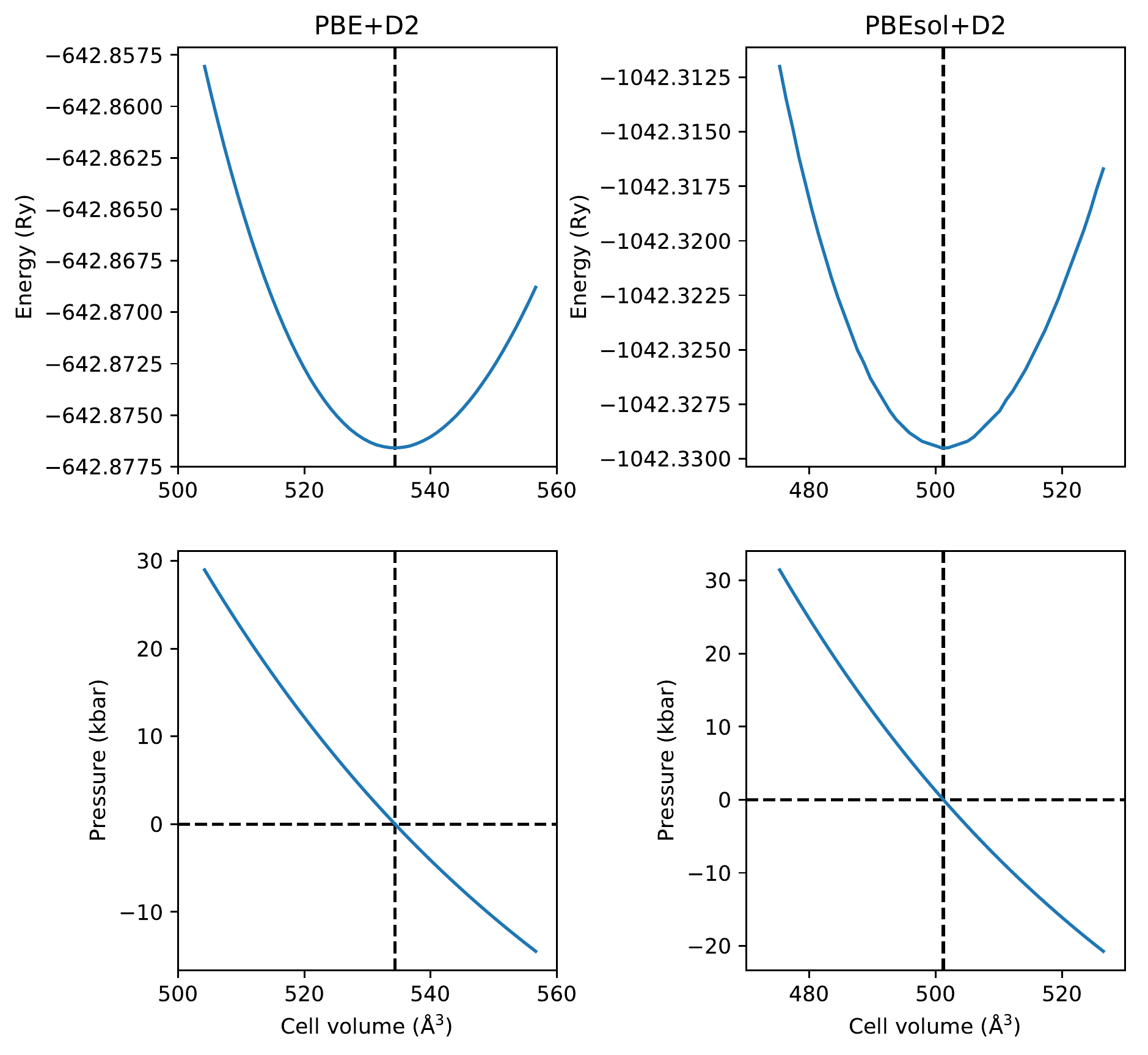}
\caption{Equation of state of PEDOT. (left) PBE+D2; (right) PBEsol+D2.
The vertical dashed line indicates the equilibrium volume.}
\label{fig:eos}
\end{center}
\end{figure}

After obtaining the optimized lattice parameters $a$, $b$, $c$ and
$\beta$, we calculate the equation of state of PEDOT by computing the
energy versus volume curve. At its minimum energy, the equilibrium volume
is 534.36~\AA$^3$ for PBE and 501.22~\AA$^3$ for PBEsol, both including
the Grimme+D2 correction~\cite{Grimme2011}. The results are shown in
Fig.~\ref{fig:eos}. The good convergence (cutoff and k-points sampling)
of our calculations is supported by the fact that the calculated pressure
(from the trace of stress tensor) vanishes at equilibrium.

\begin{figure}
\begin{center}
\includegraphics[width=\columnwidth]{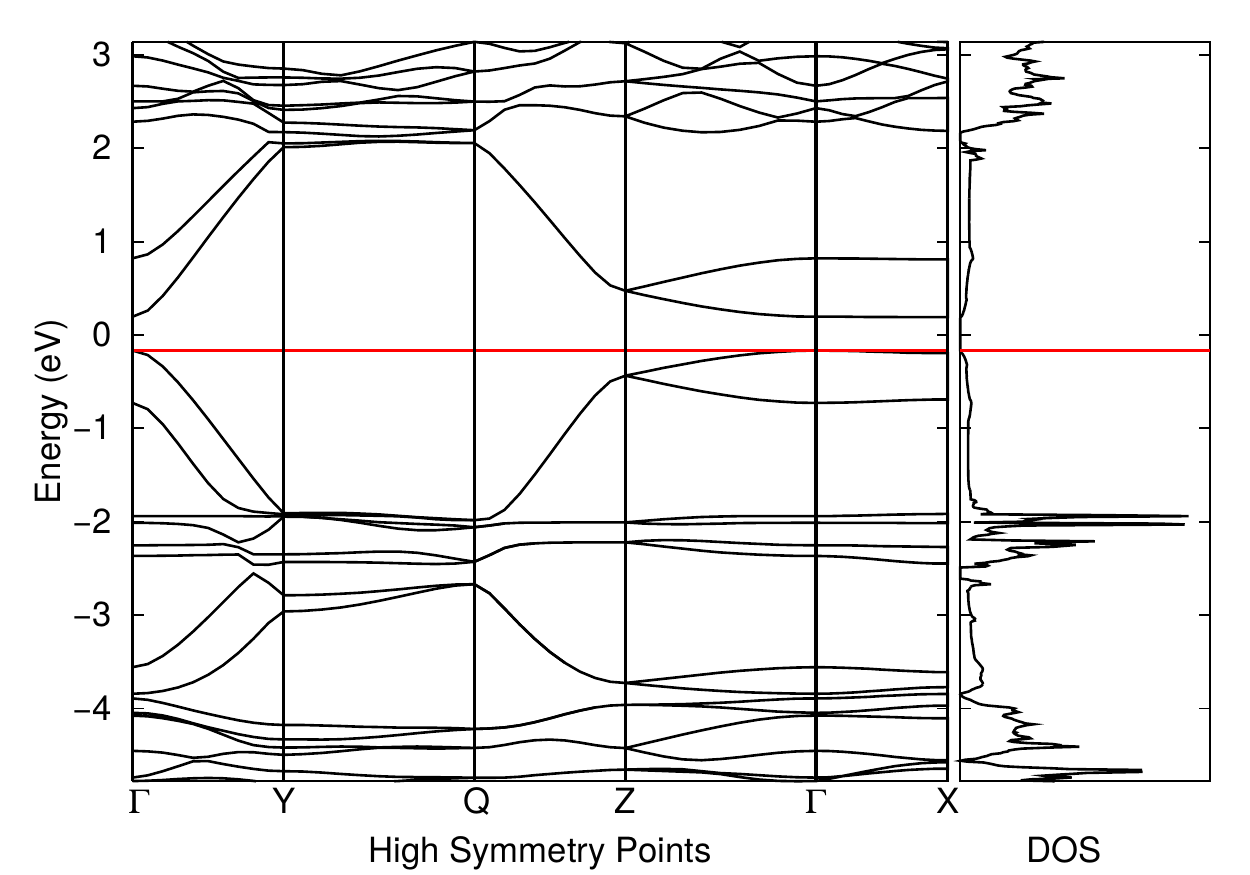}
\includegraphics[width=\columnwidth]{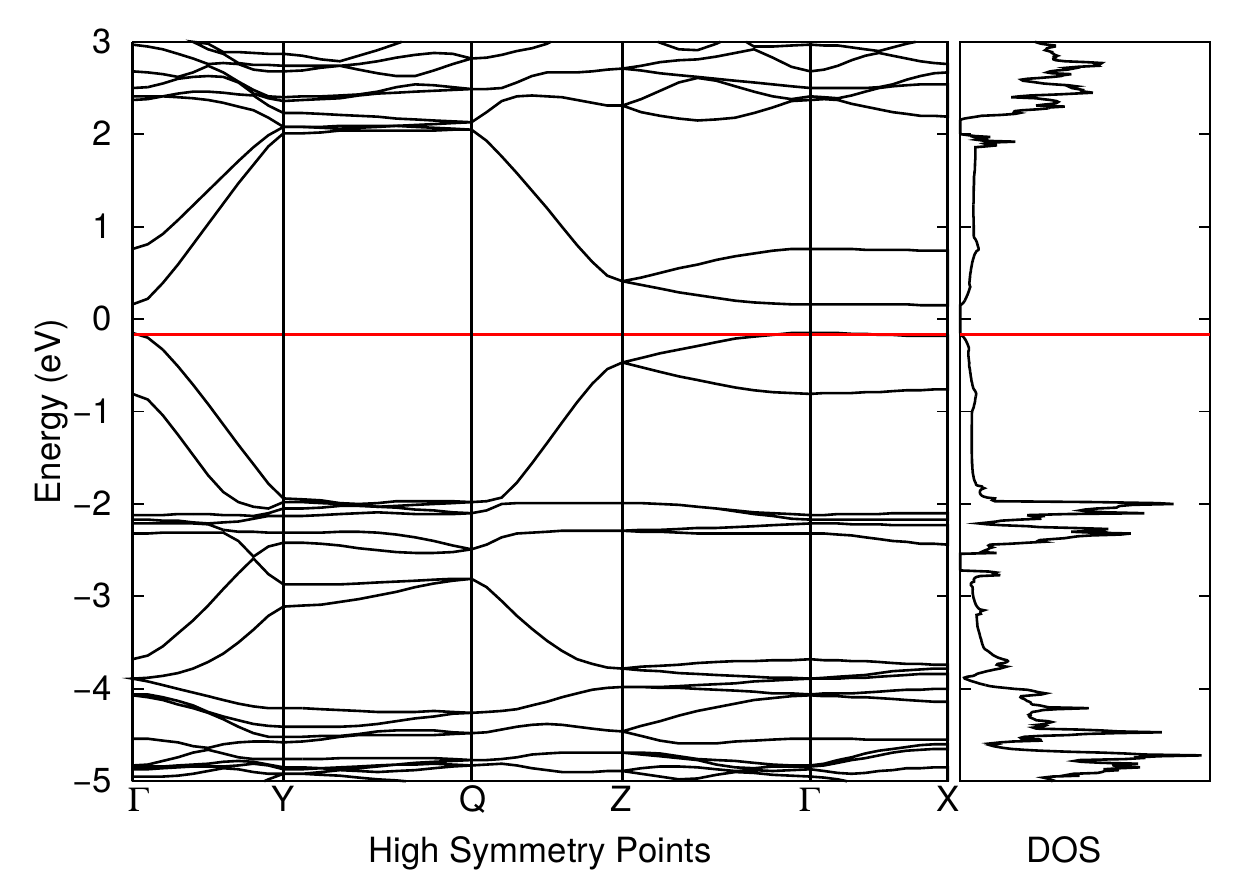}
\caption{Band structure and density of states at equilibrium volume,
obtained with PBE (top panel) and PBEsol (bottom panel). The horizontal
red line indicates the top of the valence band.}
\label{fig:bands}
\end{center}
\end{figure}

We calculated direct band gaps of 0.32~eV and 0.42~eV respectively in
PBEsol and PBE exchange-correlation functionals. In both cases, flat-bands
occur between 2 to 3~eV and between $-$4 to $-$5~eV, which are due to the
absence of electron hopping at those energies, and manifest as sharp
peaks in the density of state plot. The density of states at the valence
and conduction band edges is quite small, and the band structure displays
an oval-shaped band from $\Gamma$ to $Z$, with nearly symmetric valence
and conduction dispersion. The bands also show significant band gaps
at the $Y$ and $Q$ high symmetry points. The band structures are similar
to what was reported in Refs.~\cite{Shi2015,Kim2008,Zhang2015}. In this
work, PBE exchange-correlation functional produces a larger band gap. This
variation in band gap is due to the difference in the exchange-correlation
term supplied to the Kohn-Sham equation within DFT. Although there is no
report on the experimental band gap of bulk pristine PEDOT, from the study
of Ref.~\cite{Charles2016}, PBEsol predicted a small band gap from DFT
study of some oxyfluoride compounds. The calculated PBE and PBEsol-PAW
band gaps in this work are in good agreements with 0.45~eV reported
in Ref.~\cite{Zhang2015} and 0.37~eV computed by Kim~\cite{Kim2008},
although far from 0.53~eV and 0.16~eV reported in Ref.~\cite{Shi2015}.

DFT is in principle exact, the issue is with the exchange-correlation (XC)
functionals. It is a well known fact that local and semilocal XCs (such as LDA
and GGAs) severely underestimates band gaps by a factor $\sim$~2 in most systems,
but from Tab.~\ref{tab:gap} it is apparent that the underestimation is much more
dramatic in the present case. Hybrid-DFT functionals (such as HSE06) are known
to mitigate this problem and yield gaps larger that the PBE family. In fact,
the largest band gap in Tab.~\ref{tab:gap} has been obtained with HSE06. However,
from Tab.~\ref{tab:lattice}, the crystal structure obtained by Shi
and co-workers, overestimates the lattice spacing (hence the volume)
of the crystal, and this results in a band gap reduction. For a
reliable estimate of crystalline PEDOT band gap, one could perform a
GW-BSE~\cite{Tiago2005} calculation, on the equilibrium volume that
is found by PBE and PBEsol. Indeed, Ref.~\cite{Ferretti2012}, report
that the GW-BSE band gap of several conjugated polymers is roughly 
2.2--2.5 larger than that of PBE.

\begin{table*}
\begin{center}\begin{tabular}{lllll}
\hline\hline
Method & Type of PEDOT & Cryst. structure & Band gap (eV) & Ref. \\
\hline
PW-PP (PBE+D2)     & undoped & monoclinic & 0.42 (direct) & This work \\
PW-PAW (PBEsol+D2) & undoped & monoclinic & 0.32 (direct) & This work \\
PW-PP (BLYP)       & undoped & othorhombic & 0.37 (direct) & \cite{Kim2008} \\
PW-USPP (PE)       & undoped & othorhombic & 0.45 (direct) & \cite{Zhang2015} \\
PW-PAW (PBE+D2)    & undoped & monoclinic & 0.16 (direct) & \cite{Shi2015} \\
PW-PAW (HSE06)     & undoped & monoclinic & 0.53 (direct) & \cite{Shi2015} \\
\hline
Expt. (optical spectroscopy) & thin film & & 1.5--1.6 & \cite{Pei1994} \\
Expt. (vis-IR abs. spectra) & film & & 1.64 & {}\cite{Havinga1996} \\
Expt. (spectroscopy, electrochemistry) & thin film & orthorhombic & $\sim$1.7 & \cite{TranVan2001} \\
\hline\hline
\end{tabular}\end{center}
\caption{Comparison of band gap obtained in this work, previous calculations
and experiments.}
\label{tab:gap}
\end{table*}

\begin{figure}
\begin{center}
\includegraphics[width=\columnwidth]{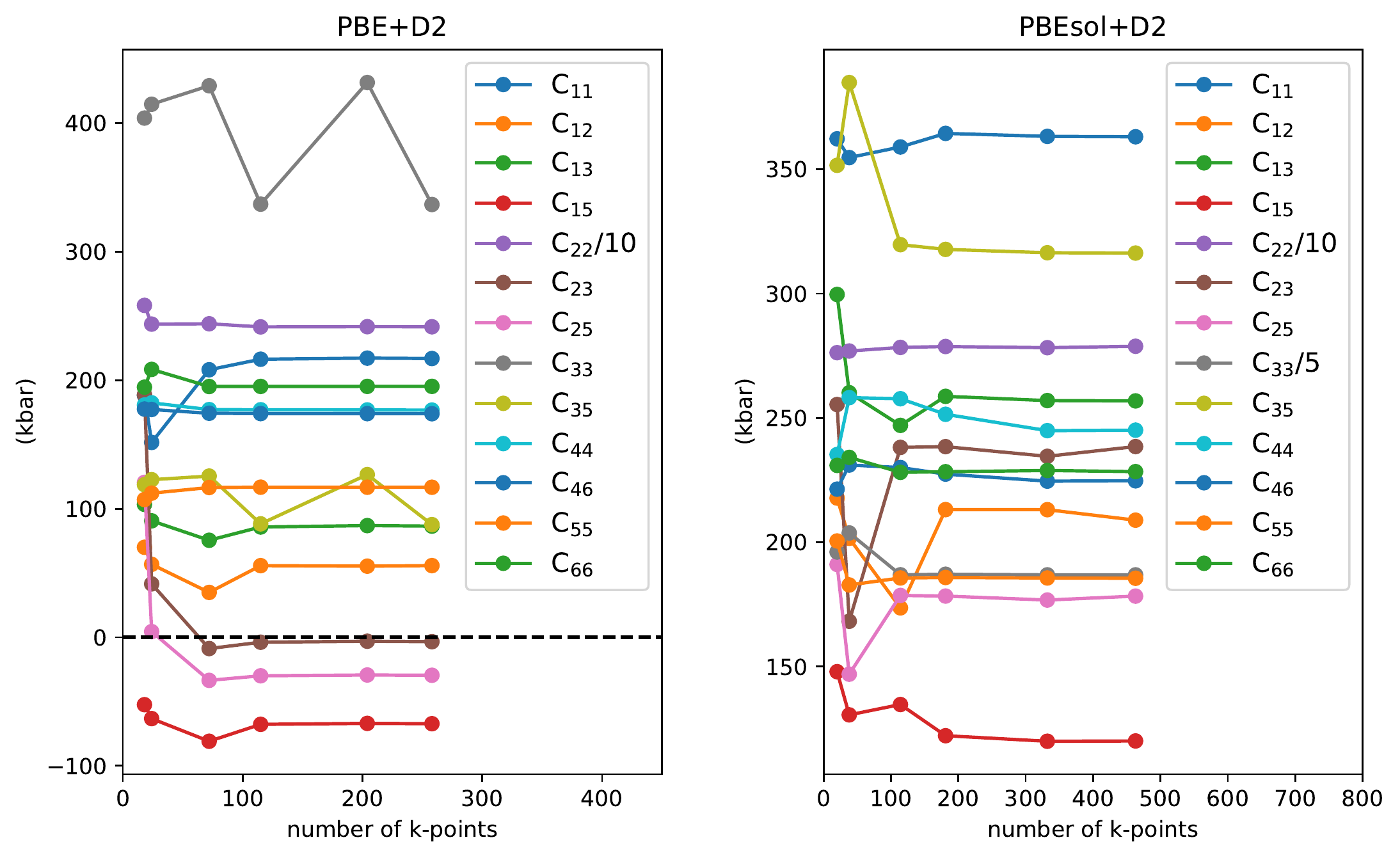}
\caption{Calculated elastic constants as a function of k-point
sampling. The $C_{22}$ and $C_{33}$ have been rescaled for sake of
clarity.}
\label{fig:elastic}
\end{center}
\end{figure}

\begin{table*}
\begin{center}\begin{tabular}{llllllllllllll}
\hline\hline
Method & C$_{11}$ & C$_{12}$ & C$_{13}$ & C$_{15}$ & C$_{22}$ & C$_{23}$ & C$_{25}$ & C$_{33}$ & C$_{35}$ & C$_{44}$ & C$_{46}$ & C$_{55}$ & C$_{66}$ \\
\hline
PBE+D2    & 216.9 & 55.8 & 86.5 & $-$67.3 & 2416.1 & $-$3.4 & $-$29.6 & 336.7 & 87.7 & 176.8 & 173.9 & 116.7 & 95.2 \\
PBEsol+D2 & 363.1 & 208.9 & 256.9 & 120.1 & 2788.2 & 238.5 & 178.3 & 747.6 & 361.3 & 245.0 & 224.7 & 185.5 & 228.4 \\
\hline\hline
\end{tabular}\end{center}
\caption{The 13 independent elastic constants, calculated using the PBE and PBEsol
XC functionals}
\label{tab:elastic}
\end{table*}

We calculated the converged 13 independent elastic constants,
with respect to k-point sampling (Fig.~\ref{fig:elastic} and
Tab.~\ref{tab:elastic}). The difference between PBEsol and PBE are mainly
related to the small differences in the lattice spacing. Moreover, it
is a well-known fact that PBE tends to underestimate crystal cohesive
energy and elastic moduli. In the present case, PBE yields negative
off-diagonal elastic constants. A negative elastic constant is not
forbidden, as long as it fulfills all the Cauchy relations for mechanical
stability~\cite{Mouhat2014}.  Indeed, both PBE and PBEsol elastic constant
matrices have positive eigenvalues, which is the necessary and sufficient
condition for the mechanical stability. Our results show a similar trend
to the ones reported in Ref.~\cite{Peng2015}, calculated for the
$\beta$-cyclotetramethylene tetranitramine molecular crystal.

\begin{figure}
\begin{center}
\includegraphics[width=\columnwidth]{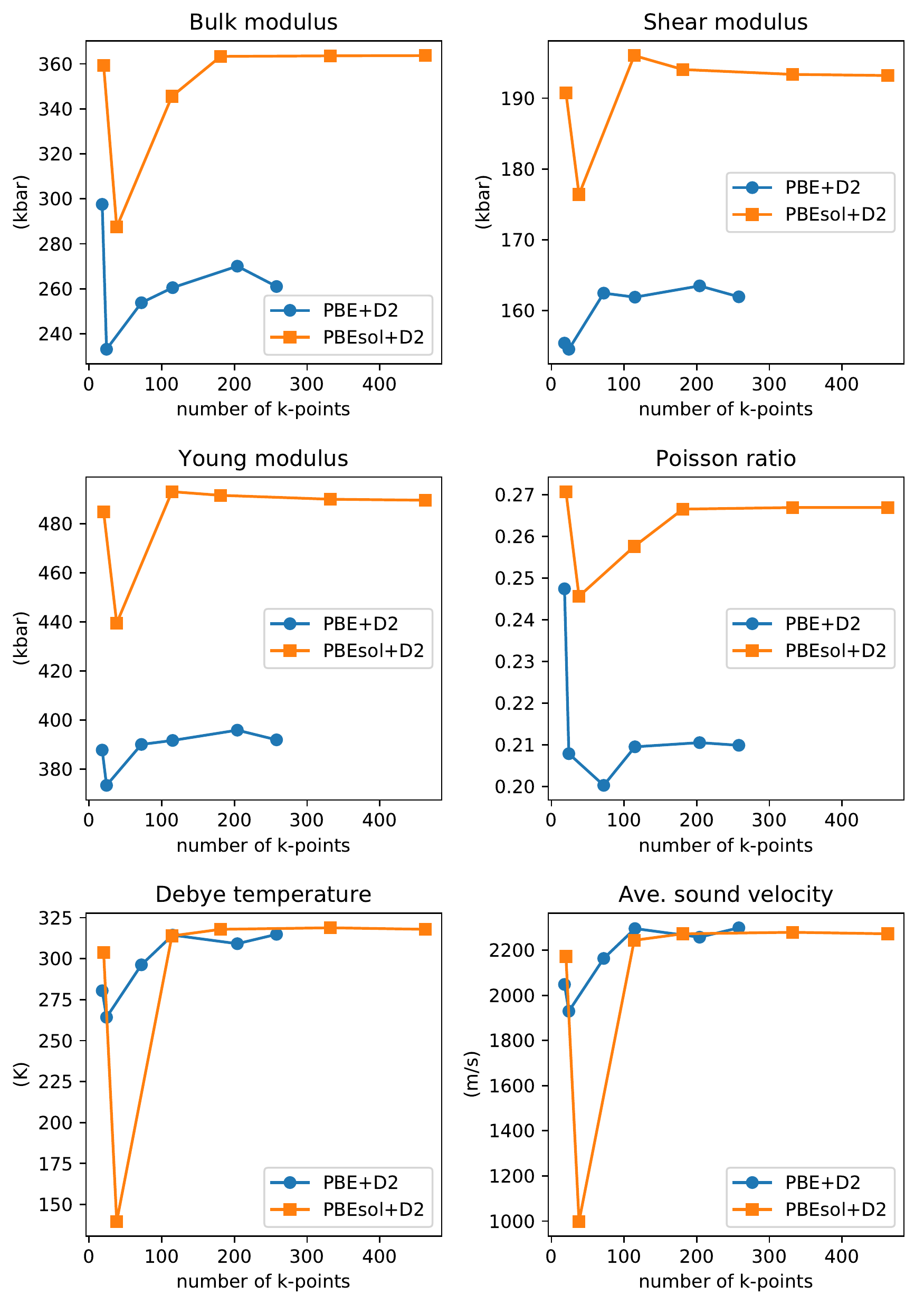}
\caption{Calculated elastic properties as a function of k-point sampling.}
\label{fig:modulii}
\end{center}
\end{figure}

\begin{table*}
\begin{center}\begin{tabular}{lllllllll}
\hline\hline
Method & Bulk    & Shear   & Young   & Poisson & Pugh  & Vicker's & Debye       & Ave. speed \\
       & modulus & modulus & modulus & ratio   & ratio & hardness & temperature & of sound \\
       & (kbar)  & (kbar)  & (kbar)  &         &       & (kbar)   & (K)         & (m/s) \\
\hline
PBE+D2    & 261.0 & 161.0 & 391.9 & 0.2099 & 0.6203 & 19.4 & 315 & 2299.1 \\
PBEsol+D2 & 363.7 & 193.2 & 489.5 & 0.2669 & 0.5312 & 17.7 & 318 & 2272.5 \\
\hline\hline
\end{tabular}\end{center}
\caption{Computed Hill elastic modulii of crystalline PEDOT.}
\label{tab:moduli}
\end{table*}

From the elastic constants, we extracted the averaged Hill elastic modulii
and as well as the Poisson ratio~\cite{Hill1952}. Our results, highlight the
importance of k-point sampling in computing elastic properties. In
fact, while a relatively coarse k-point sampling is sufficient to
converge the total energy, converging derivatives of the total energy
(forces and stress) usually require finer k-point meshes.  With PBEsol,
we reach convergence after 463 k-points resulting in a maximum 0.08\%
difference in the moduli with respect to 332 k-points. Similarly, with
PBE and using norm-conserving pseudopotentials, 258 k-points yield 1-3\%
variation in the moduli, compared to 204 k-points. Due to the large
number of k-point, these calculations required a substantial computational
effort.  The Pugh ratio shows a disagreement on its brittleness as PBEsol
categorize pristine PEDOT as ductile while PBE calculations predict a
brittle material with a greater Vickers hardness. With a hardness of
1.77~kbar and 1.94~kbar, PEDOT is far from that of diamond which is a
super-hard material but has its hardness similar to that of ZnS with
experimental Vicker's Hardness of 18~kbar~\cite{Tian2012,Chen2011}. Since
the ratio of Bulk Modulus to Share Modulus is greater than unity, and
the Poisson ratio tends weakly towards 0.5, we predict pristine PEDOT
as mildly incompressible.

In Fig.~\ref{fig:elastic}, one can note that the $C_{33}$ and $C_{35}$
elastic constants are not fully converged with respect to k-points sampling.
In principle it is possible to calculate the error propagation from
the elastic constants, to the Hill elastic modulii. However, to have an
estimate, we computed the change of elastic modulii due to a variation
of the $C_{33}$ (from 336 to 420 kbar) and $C_{35}$ (from 87 to 100 kbar)
elastic constants. We found that these variations have a small effect
on the Hill modulii. The maximum variation of the bulk, Young and
shear modulus is 6, 10 and 4 kbar, respectively. The Poisson ration
remains basically unchanged. Therefore, the large difference between
the PBE+D2 and PBEsol+D2 calculated elastic modulii can be explained by
the large difference between the elastic constants, and not by the uncertainty
of the individual elastic constants

\begin{figure}
\begin{center}
\includegraphics[width=\columnwidth]{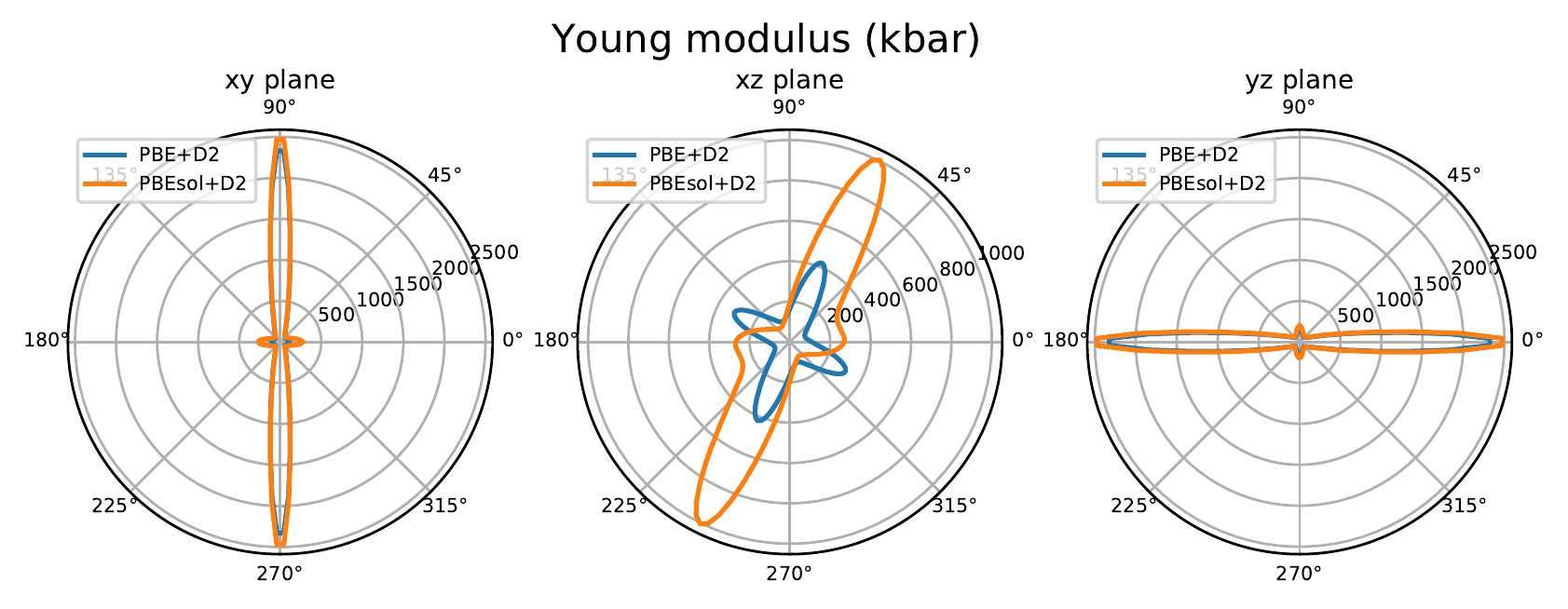}
\includegraphics[width=\columnwidth]{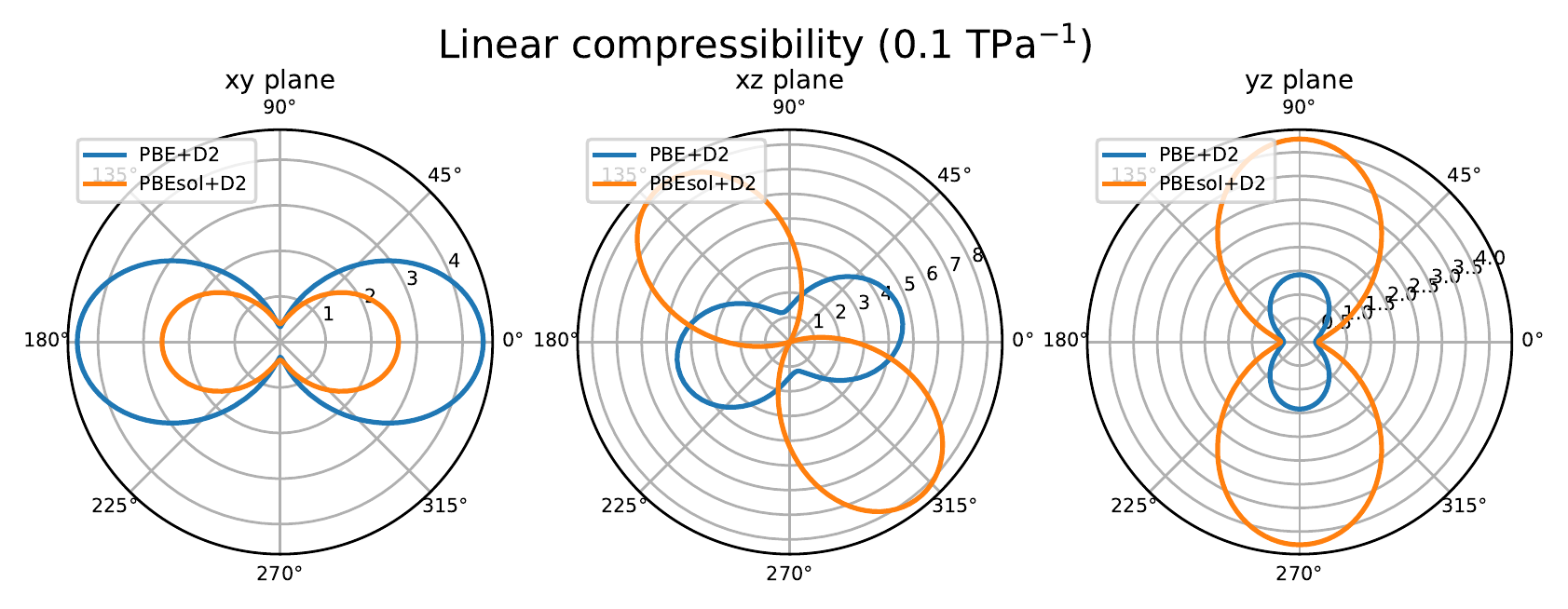}
\includegraphics[width=\columnwidth]{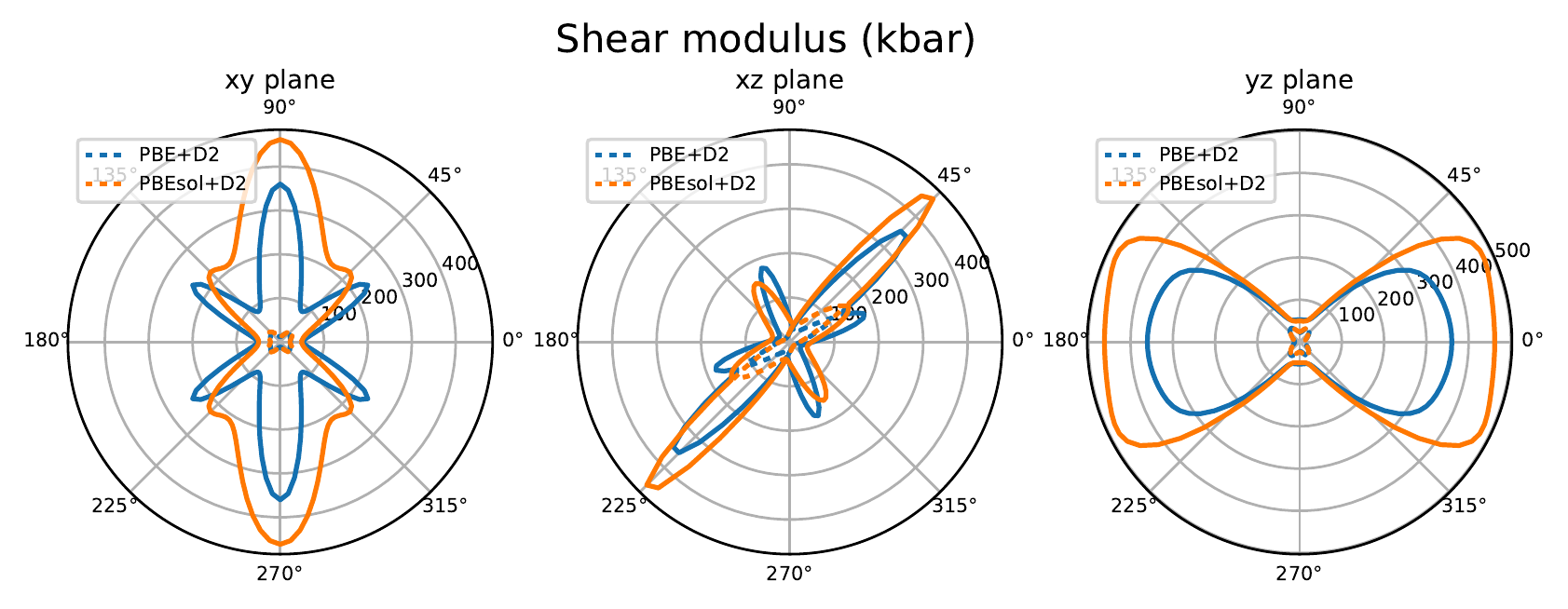}
\caption{Calculated directional elastic modulii. Solid lines correspond to
maximum values. Dotted lines to mininum values. The polymeric chains are
parallel to the $y$ axis.}
\label{fig:directional}
\end{center}
\end{figure}

For highly anisotropic materials, it is instructive to report the
directional elastic modulii (see Fig.~\ref{fig:directional}). The Young
modulus is largest in the $y$ direction, parallel to the polymer chain,
both for PBE+D2 and PBEsol+D2. However, the linear compressibility behaves
differently. It is largest for PBEsol+D2 in the $xz$ plane, with an angle
of $-$45$^\circ$ with the $x$ axis. This means that in PBEsol+D2 it's
easier to compress the polymer \emph{lamella} along in the $\pi$-$\pi$
stacking direction, and to reduce the inter-chain distance (see also
Fig.~\ref{fig:cell}).  Conversely, the PBE+D2 functional predict that
PEDOT is more compressible along the $x$ direction, such to reduce the
distance between chains which are co-planar. Finally, the Shear modulus
behaves similarly with the two XC functionals. Mechanical experiments
on highly crystalline, fiber oriented PEDOT, might help to ascertain
which functional is better in describing the elastic properties of PEDOT.

\section{Conclusions}\label{sec:conclusions}
This paper presents a study of ground state properties, elastic and
thermodynamic properties of crystalline PEDOT. Our results are in
relatively good agreement with experimental data. The mismatch in the Pugh
ratio that results in classifying pristine PEDOT as ductile and brittle
with different functionals and pseudopotentials is open to debate. We
predicted pristine PEDOT is mildly incompressible. Our results provide
new physical insights to the elastic and mechanical properties of PEDOT.

\section*{Acknowledgments}
The authors acknowledge the Abdus Salam International Centre for
Theoretical Physics for computational access to its Clusters, ROA and GA
are grateful to Ivan Girotto for assistance on resolving computational
bugs. DC acknowledges Alberto Bossi (CNR-ISTM) and Donato Belmonte
(University of Genova) for useful discussions.

\bibliography{PEDOT}

\end{document}